\documentclass[5p]{elsarticle}

\usepackage{amsmath}
\usepackage{booktabs}
\usepackage{graphicx}
\usepackage{hyperref}
\usepackage{cleveref}
\usepackage{listings}
\usepackage{xcolor}
\usepackage{url}
\usepackage{pifont}
\crefname{lstlisting}{listing}{listings}
\Crefname{lstlisting}{Listing}{Listings}

\definecolor{yamlkey}{RGB}{0,0,180}
\definecolor{yamlvalue}{RGB}{128,0,0}
\definecolor{yamlliteral}{RGB}{0,128,0}

\lstdefinelanguage{yaml}{
  basicstyle=\ttfamily\footnotesize,
  keywordstyle=\color{yamlkey},
  commentstyle=\color{gray}\itshape,
  stringstyle=\color{yamlvalue},
  morecomment=[l]{\#},
  moredelim=[s][\color{yamlliteral}]{:}{\ },
  sensitive=false,
  showstringspaces=false,
  breaklines=true,
  breakatwhitespace=false,
  columns=fullflexible,
  frame=single,
  framesep=2mm,
  rulecolor=\color{black!30},
  backgroundcolor=\color{black!5},
  xleftmargin=2mm,
  xrightmargin=2mm
}

\journal{DFRWS USA 2026}
\biboptions{authoryear}

\newcommand{\adare}{ADARE}
\newcommand{\adareclient}{ADARE-Client}
\newcommand{\adareserver}{ADARE-Webapp}
\newcommand{\fortracepp}{ForTrace\texttt{++}}
\urldef{\adaredocsurl}\url{https://fkie-cad.github.io/adare/}
\urldef{\adaredemourl}\url{https://fkie-cad.github.io/adare/paper/demos.html}
\urldef{\adareclienturl}\url{https://github.com/fkie-cad/adare}
\urldef{\adareserverurl}\url{https://github.com/fkie-cad/adare-server}
\urldef{\playbookexample}\url{https://github.com/fkie-cad/adare/blob/main/paper/demo}
\urldef{\experimentsurl}\url{https://github.com/fkie-cad/adare/tree/main/paper/experiments}

\newcommand{\ncasestudies}{five}
\newcommand{\statetransitiontesting}{State Transition Testing}

\begin{document}

\begin{frontmatter}
\title{Do you dare to try Test-Driven Forensics? Increasing Trust in Desktop Forensics with ADARE}

\author{Michael Külper\corref{cor1}}
\ead{michael.kuelper@fkie.fraunhofer.de}
\author{Martin Lambertz}
\ead{martin.lambertz@fkie.fraunhofer.de}
\author{Mariia Rybalka}
\ead{mariia.rybalka@fkie.fraunhofer.de}
\address{Fraunhofer FKIE, Zanderstr. 5, 53177 Bonn, Germany}
\cortext[cor1]{Corresponding author.}

\begin{abstract}
Digital forensic relies on validated tools and established procedures, yet the underlying operating systems, applications, and analysis tools evolve rapidly. This evolution can cause artifact behavior and tool outputs to drift, silently degrading repeatability and confidence in long-lived forensic interpretations. We present test-driven forensics, a practical approach that treats forensic expectations as executable specifications: expected artifacts and expected tool outputs are encoded as tests that can be rerun across versions to detect regressions. Crucially, our approach also enables \statetransitiontesting{}, validating the system’s expected state after each user action rather than only performing post-mortem checks on a final disk image; this supports causal attribution and makes transient behavior testable.
We implement the methodology in \adare{}, an open-source framework that runs controlled experiments in virtual machines and simulates realistic user activity via computer-vision-guided GUI automation. \adare{} includes a companion web platform for sharing experiments, environments, and results to facilitate independent reruns and peer verification. We evaluate \adare{} in five case studies spanning artifact research and tool validation. In particular, a 25-version regression study of Autopsy reveals substantial, largely undocumented changes in exported report outputs, demonstrating how executable tests make drift measurable and reproducible at scale.

\end{abstract}


\end{frontmatter}

\section{Introduction}\label{sec:intro}
\sloppypar
Digital forensics demands adherence to \textit{``established principles, standards, and processes''}~\citep{Arnes/DF/2-DF-Process}, yet the discipline faces persistent challenges in meeting this ideal. 
While foundational definitions emphasize the use of validated tools and scientifically derived methods~\citep{Palmer-2001-dfrws-roadmap}, current practice often lags behind due to the rapid evolution of software. The process of identifying artifacts remains largely manual and is frequently documented in informal sources, such as blog posts, which are unsuitable for legal proceedings. 
This problem is amplified by the dynamic nature of modern systems: artifact structures and meanings often change with minor software updates, rendering prior knowledge obsolete or incorrect~\citep{horsman2019raiders, singh2017program, amcache-blog-post, lagny-amache}.

To address these challenges, we propose test-driven forensics: a methodology in which expected artifact behaviour and forensic tool outputs are encoded as executable tests that can be rerun across OS and application versions to establish a verifiable reference baseline. We implement this methodology in \adare{}, a framework that automates the execution of controlled experiments and checks expected artifacts and tool results across versions. We focus on desktop operating systems (OS), where frequent OS and application updates change artifacts and frequent tool updates can introduce tool-output drift. By adapting established software engineering practices---such as automated unit, system, and regression testing---to the forensic domain, we enable continuous validation and regression testing of artifact expectations and tool outputs as systems evolve.

\adare{} provides the technical infrastructure to support these practices by simulating realistic user behavior through computer-vision-guided GUI automation. Pairing a simulated user action with a test of its expected outcome is analogous to automated testing in software engineering~\citep{Cohn2010SucceedingWithAgile, Martin2011CleanCoder}. 
This enables standardized experiments that measure artifact behavior and tool-output drift over time.

To foster collaboration, the framework includes a web platform for sharing experiments and their results. This allows researchers to replicate findings with minimal setup, strengthening confidence through independent validation. Using a shared experiment repository and rerunnable workflows, \adare{} supports a community process for validating artifacts and tool behavior that are beyond the scope of any single organization to maintain~\citep{McKemmish-1999-Forensic-Computing, craiger2006le-and-di}.

This work makes the following scientific contributions:
\begin{itemize}
\item \textbf{Test-Driven Methodology:} We introduce a methodology that treats artifact knowledge as executable tests, enabling continuous validation and regression testing of forensic tools and artifact behavior across OS and application versions.
\item \textbf{\adare{} Framework \& Ecosystem:} We present an open-source system that uses computer-vision-guided GUI automation to execute self-validating experiments, with a web platform for sharing and reproduction.
\item \textbf{Evaluation:} We demonstrate the approach through \ncasestudies{} case studies that expose version-specific artifact divergence and tool-output drift.
\end{itemize}

\section{Related Work}\label{sec:related-work}

\subsection{Artifact Research}
Digital forensic artifact research is fundamentally driven by the escalating volume, variety, and complexity of digital evidence, with the core aim of identifying and reliably interpreting information that is of forensic utility. A central challenge is reducing missed or misinterpreted evidence through standardized methods and broadly accepted interpretations. To systematically manage this growing body of knowledge, researchers have developed centralized, crowdsourced repositories, such as the Artifact Genome Project~\citep{agp-grajeda-2018}, the Artifact Catalog~\citep{casey2022crowdsourcing}, the Digital Forensics Artifacts Repository~\citep{df-artifact-repo}, the DFIR Artifact Museum~\citep{dfir-artifact-museum}, and the knowledge base SOLVE-IT~\citep{hargreaves2025solve}, which archive vetted artifacts in a structured and searchable form.
These catalogs formalize artifact representation by detailing necessary and sufficient elements, including associated context, application, platform, and extraction recipes. The accurate interpretation of collected traces depends significantly on reference data, defined as the standard information used to classify or assign meaning to data acquired from digital sources~\citep{spichinger2025evidencemeaning}. However, highly dynamic and evolving systems pose a challenge, as reference data must be gathered promptly to prevent the decay of the assigned meaning of preserved traces over time, increasing overall uncertainty. Methodological frameworks, such as the Framework for Reliable Experimental Design (FRED)~\citep{Horsman-FRED}, guide the robust planning, implementation, and analysis phases of artifact testing to ensure reliable, repeatable, and factually accurate results suitable for legal admissibility. Furthermore, specialized tools like Argus~\citep{argus-aardwolf-2025} facilitate dynamic analysis by monitoring the file system during controlled application usage experiments, storing these results in reference databases like Aardwolf.

The above approaches underscore the necessity of a centralized repository for artifact knowledge, yet they also reveal the considerable challenges inherent in its creation and long-term maintenance. Related foundational work addresses complementary concerns: FAIR-aligned reproducibility workflows~\citep{moreau2023containers} target experiment sharing through containerization,~\cite{carrier2006hypothesis} models file system state, and~\cite{garfinkel2012general} propose systematic image comparison via differential analysis. Building upon these efforts, this work shifts from post-hoc image comparison to in-run, assertion-based validation of artifact behavior and tool outputs across versions.

\subsection{Tool Testing}\label{sec:rel-work-tool-testing}
Based on a review of various publications, we extracted several key aspects of effective digital forensics tool testing. The process is predicated on rigorous validation---confirming a tool meets requirements for its intended use---and subsequent verification, which confirms this validation within a specific laboratory's unique environment~\citep{horsman2018yourhonour,flandrin2014dfts,guo2009validation}. 
This dual process is foundational to satisfying the stringent demands of legal admissibility standards and for achieving laboratory accreditation under frameworks like ISO 17025~\citep{guo2009validation,wilsdon2006blackbox}. Given the proprietary, closed-source nature of most commercial software, the predominant methodology is a functionality-oriented, black-box testing paradigm, which is more feasible than approaches requiring source code access~\citep{wilsdon2006blackbox,10yrs-of-cftt-2011}. Here, a structured testing process commences with a formal plan that identifies a tool's discrete functions, for which detailed test cases and documented reference datasets are created~\citep{craiger2006validation-of-df-tools,bhat2021trusted,Brunty2023,10yrs-of-cftt-2011}. During execution in a controlled environment, the tool's output is systematically compared against a pre-defined result acceptance spectrum to determine if it functions as specified. Critical to this evaluation is recognizing that errors in forensic tools are typically systematic, not statistical; consequently, testing must focus on thoroughly documenting failure types and the specific conditions that trigger them rather than pursuing a single error rate~\citep{lyle2010errorrate}. 
Recent work has expanded on this by proposing abstract models for systematic error mitigation~\citep{hargreaves2024abstract} and scenario-based quality assessments for specific tool categories~\citep{rzepka2025scenario}.
Furthermore, advanced testing should incorporate adversarial scenarios to assess a tool's resilience against anti-forensic techniques~\citep{bhat2021trusted,wundram2013antiforensics}. 
While performance testing is also significant for tool testing in digital forensics~\citep{pan2009performance,SWGDE2024testing}, this work focuses solely on the correctness of the tools.

\subsection{Data Synthesis Frameworks}
The lack of availability of datasets is a known problem in digital forensics~\citep{grajeda2017datasets}. In response, various data synthesis frameworks have been developed to generate forensically relevant data. Since these frameworks often function by emulating user interaction, they are viable solutions for the automated generation of digital artifacts required for forensic research and tool validation.

\cite{goebel2022fortrace}, \cite{schmidt2023trace-synthesis}, and \cite{wolf2024fortracepp} conducted surveys of such data synthesis frameworks, assessing their respective capabilities for generating realistic forensic test data. These reviews collectively analyze a range of existing solutions, including Forensig$^{2}$~\citep{moch2009forensig}, ForGeOSI\footnote{\url{https://github.com/maxfragg/forgeosi}}, ForGen\footnote{\url{https://github.com/Jjk422/ForGen}}, EvilPlant~\citep{scanlon2017eviplant}, VMPOP~\citep{park2018vmpop}, hystck~\citep{goebel2020hystck}, TraceGen~\citep{du2021tracegen}, ForTrace~\citep{goebel2022fortrace}, and \fortracepp{}~\citep{wolf2024fortracepp}.

These data synthesis frameworks are frequently developed to address the need for test datasets in digital forensics, particularly for practitioner training, educational purposes, and machine learning applications. 
A primary objective of these frameworks is to generate datasets with high fidelity by emulating the digital traces that result from authentic human-computer interactions. 
This process aims to minimize the introduction of artifacts that would reveal the automated nature of the data creation. 
The absence of such automation artifacts is relevant for creating realistic scenarios, however, their presence may be less consequential for other applications, such as targeted tool and method validation. 
In these validation scenarios, the ground truth is comprehensively known, which allows an examiner to account for and disregard any artifacts known to originate from the data generation process.

While the primary goals of the surveyed data synthesis frameworks may differ from our use case, the underlying research provides critical insights relevant to this study. \cite{du2021tracegen} indicate that the nature and volume of the generated artifacts are contingent upon the specific automation method employed. To achieve higher forensic fidelity in the test images, the authors recommend an approach that utilizes GUI automation and computer vision, as opposed to relying on simpler, and potentially less realistic, command-line or direct script execution methods---a finding confirmed by \cite{schmidt2023trace-synthesis}.

Moreover, \cite{goebel2022fortrace} and \cite{wolf2024fortracepp}  emphasize the necessity of capturing a comprehensive set of digital traces, including not only artifacts on persistent storage but also data from volatile memory and network traffic. A prevailing architectural approach in recent frameworks is the agent-less utilization of hypervisor functionalities, combined with computer vision, to simulate authentic user behavior.
The implications of this design choice for test-driven forensic experimentation are considered in the design of our methodology.

\section{A Test-Driven Methodology for Artifact Research and Tool Validation}\label{sec:methodology}
Drawing on established software engineering paradigms, we propose a methodology that adapts testing practices to two forensic areas: artifact research and tool testing. 
The methodology relies on two core components:
\begin{itemize}
    \itemsep0pt
    \item[(1)] the automated simulation of user behavior, and
    \item[(2)] tests applied to validate the resulting system state.
\end{itemize}

In the following, we refer to a combination of simulated user behavior and one or more corresponding tests as an \emph{experiment}. 
The system in which these experiments are executed is called the \emph{environment}.

\subsection{Simulating User Behavior}
Within this methodology, we describe user simulation not merely as a state transition, but as \emph{constructive simulation}---the use of computational agents to replicate the interaction modalities of human users. 
The objective of this simulation varies depending on the specific forensic challenge being addressed. 
In the context of tool validation, the ``user'' corresponds to the forensic investigator operating the analysis software. 
Conversely, in artifact research, the term refers to a standard end-user performing interactions within an operating system or software application.
In both scenarios, simulating user behavior must preserve the \emph{realism} required by the testing objectives. 

\paragraph{Realism}
To ensure forensic validity, this methodology a\-dopts the formal definition of realism based on the ``Prover/Verifier'' model proposed by \cite{voigt2025metrics}. 
In this model, synthetic data is considered realistic if it is indistinguishable from real data when analyzing a specific set of ``Allowed Features'' ($A \subset F$, where $F$ denotes all measurable forensic features). 

Within our methodology, we define $A$ to encompass all target forensic artifacts and tool outputs generated by the simulated user, while explicitly excluding traces inherent to the automation infrastructure itself---such as hypervisor logs or agent remnants---provided these do not obfuscate the target evidence. 
This distinction imposes a strict constraint on the simulation method: because exchanging GUI interactions with low-level scripting frequently generates metadata that deviates from genuine human interaction and effectively corrupts the features within $A$ \citep{du2021tracegen}, our methodology requires the use of authentic GUI simulation to ensure the resulting forensic artifacts are indistinguishable from those created by a human user.

\subsection{Testing Strategies}\label{subsec:testingstrategies}
Beyond user simulation, the second part of the methodology is testing the resulting system-state changes against an expected outcome.
In forensic tool validation, this means determining whether a generated output, exemplarily a report, accurately reflects the activity inferred from a given artifact source.
Conversely, in artifact research, testing focuses on confirming whether specific artifacts are created or modified in response to simulated actions.

The design of experiments depends on the study objectives. This includes deciding how detailed the tests should be and whether to separate simulated user actions from the tests or combine them.
We therefore present several possible testing strategies derived from established software testing principles.

\paragraph{Test Pyramid}
Based on the test pyramid, we can organize experiments into three distinct levels of granularity:
\begin{itemize}
    \item \textbf{Unit Level:} At the most granular level, experiments target atomic aspects. For artifact research, this involves validating individual characteristics of an \textit{atomic artifact} (cf. \cite{casey2022crowdsourcing}), such as the presence of a specific log entry or a value in an SQLite database. Correspondingly, for tool testing, this involves verifying that a specific value appears correctly in the tool's parsed output.
    
    \item \textbf{Integration Level:} Ascending the pyramid, these tests address complex structures. They focus comprehensively on a single artifact container (e.g., a file or database) or a discrete functionality within a forensic tool. This level ensures that the artifact or tool functions correctly as a cohesive unit.
    
    \item \textbf{System Level:} At the uppermost UI layer, the methodology tests the complete set of artifacts generated by a sequence of user actions. This corresponds to Trace-to-Activity interpretation~\citep{horsman2025traces}, examining relationships between multiple artifact containers and verifying end-to-end tool workflows.
\end{itemize}

\paragraph{Regression Testing}
In software testing, \textit{``[r]egression testing is defined as the process of verifying that previously functioning software continues to operate correctly after modifications''}~\citep{regression-testing-engstrom}.
Regression testing plays a critical role in the digital forensics context.

First, when a patch or a new version of an analysis tool is released, it must be re-evaluated to ensure consistency with previous versions.
Reported results may change in several ways beyond traditional bugs. 
For example, formatting changes of output and exports can disrupt automated workflows, and data conversions (such as switching timestamp reporting from the original artifact’s timezone to UTC) can significantly affect interpretation.

Second, regression testing is necessary even when the analysis tool remains unchanged, as the software generating the artifacts (e.g., OS or applications) evolves.
Updates to these systems may alter the structure or behavior of artifacts, potentially invalidating prior interpretations.

\paragraph{\statetransitiontesting{}}
To validate artifact evolution, we treat the forensic target as passing through a sequence of observable states. Instead of treating experiments as monolithic blocks analyzed only after completion, we validate the expected state after each action. This approach offers two distinct advantages: first, it captures transient states, such as temporary cache files or volatile logs, that are often deleted before a final analysis. Second, it efficiently establishes causality by directly mapping user actions to artifact changes within a single continuous execution, avoiding the prohibitive resource overhead of running separate experiments for each action sequence.

\subsection{Essential Properties for Forensic Validity}
Applying these testing strategies to digital forensics imposes strict requirements on the experimental design to ensure results are forensically sound.

\paragraph{Reproducibility}
Reproducibility is essential in establishing the credibility and reliability of forensic findings, particularly when they are used in legal contexts. 
The ability to replicate results enables independent verification of conclusions, which is crucial for both scientific integrity and legal acceptance. 
Furthermore, reproducibility supports the broader scientific principle of transparency, allowing others in the community to scrutinize methods and results~\citep{pan2005reproducibility, casey2010forensic-analysis, OliveiraJr2020-promoting-df-experiments}.

We adopt the definition of reproducibility from the \cite{nap2019rr-in-science}, which requires \textit{``using the same input data; computational steps, methods, and code; and conditions of analysis''}. This definition implies that independent researchers must be able to execute the same experiment and obtain identical results. 
Consequently, achieving reproducibility necessitates that all components---including input data, the computational environment, and execution scripts---are made accessible.

\paragraph{Privacy Preservation and Legal Compliance}
A foundational requirement for a scientifically sound forensic methodology is that it operates within a robust legal and ethical framework. 
This requires the incorporation of privacy-preserving principles to ensure compliance with data protection regulations, such as the General Data Protection Regulation (GDPR), particularly when processing datasets that may contain personally identifiable information. 
In addition, a methodology must account for intellectual property considerations, including copyright, licensing terms, and any restrictions governing software components, datasets, and generated reports. 
Addressing these legal, ethical, and IP-related obligations is essential to enable responsible sharing of experimental findings within the research community~\citep{breitinger2023sharing}.

While curated datasets can mitigate certain privacy concerns, copyright and licensing restrictions pose a more fundamental challenge to reproducibility.
Experimental environments commonly include operating system components and proprietary software that are subject to restrictive licenses and therefore cannot be freely redistributed.
As a result, full experimental environments cannot always be shared directly, even when reproducibility is a core methodological objective.
Accordingly, this methodology explicitly recognizes this constraint and requires that reproducibility be supported through mechanisms that remain compliant with applicable legal and licensing restrictions, leaving the realization of such mechanisms to concrete implementations.

\paragraph{Transparent and Structured Reporting}
For digital forensic methods and tools to be admissible in court, they must meet standards of scientific soundness and credibility~\citep{sremack2007gap}. A primary method for establishing this trustworthiness is through rigorous validation and the subsequent reporting of its results~\citep{tully2020quality-standards,stoykova2023reliability,horsman2019derds,Horsman-2019-Tool-Testing}.

Achieving this standard requires that both the validation process and its outcomes are transparent and open to scrutiny. Consequently, an implementation of the methodology should support not only the reproduction of experiments but also the generation and dissemination of comprehensive reports detailing their results. 
To facilitate transparency, these reports ought to document relevant intermediate steps as well as final outcomes.

While reproducibility directly establishes the credibility of findings for the researcher conducting the replication, the broader community benefits from the public reporting of these reproduced experiments. The availability of multiple, independent reports that present consistent findings enhances the collective confidence in a specific artifact or tool. This allows other parties to place greater trust in the results, even without performing the replication themselves.

\subsection{Community-Driven Verification}
The governance of tool validation can follow either a centralized or a federated model, each with distinct advantages and disadvantages~\citep{Horsman-2019-Tool-Testing}.
A centralized model offers strict quality control and consistency but creates a bottleneck, as a single entity cannot scale to match the velocity of software updates. Conversely, a federated model offers high scalability through community contribution but risks varying standards of quality and legal compliance.

To satisfy the dual requirements of data quality and research scalability, we propose a hybrid governance model.
This model combines a centralized curation process for maintaining the quality and integrity of core experimental assets with a federated, community-driven approach to validation and peer review.

To ensure a high standard of quality, the submission of new experiments and fundamental components, such as experimental environments, is subject to a centralized review and curation process. 
This moderation is essential to maintain scientific soundness, prevent duplication, and verify the integrity of experiments and related assets.

\subsection{Methodological Applications}\label{subsec:methodological-applications}
As already stated before, the proposed test-driven methodology is designed to address challenges in two primary forensic domains: Artifact Research and Tool Testing.
Within the following we describe how this methodology advance these domains.

\paragraph{Artifact Research}\label{par:artifactresearch}
The methodology enables forensic sound explorative artifact research, one of the two research orientations identified by~\cite{Horsman-FRED} within the FRED framework. 
By automating the simulation of user behavior, the methodology enables the systematic verification of how artifacts are created or modified in response to specific simulated user actions.

This approach moves researchers from ad-hoc observation toward rigorous hypothesis testing. Expected forensic outcomes, such as specific file changes or log entries, can be formally defined and automatically evaluated. 
Unit-level tests validate atomic trace attributes, while system-level tests correlate traces across multiple artifact containers. 
In addition, \statetransitiontesting{} capabilities make it possible to map specific actions to artifact changes efficiently.

Applying regression testing principles to artifact behavior further supports the verification of consistency across operating system or software versions. 
This ensures that forensic interpretations remain reliable as software evolves and assists in automatically identifying changes in artifact structures.

Finally, the methodology addresses the current lack of transparent, reproducible testing procedures by providing a structured basis for reproducible experiments. This collaborative approach promotes peer review and helps bridge the gap between rapidly evolving software and the need for a verifiable knowledge base of forensic artifacts and their behavior.

\paragraph{Tool Testing}
In the domain of Tool Testing, the methodology replaces manual report review with automated, as\-sertion-based verification. 
This formalization of expected outcomes into executable tests offers distinct advantages across different testing scenarios. 

When applied to existing reference datasets, such as public forensic corpora, it enables continuous regression testing by encoding known ground truth into automated test suites, ensuring tool updates do not degrade parsing accuracy. 
Conversely, in scenarios requiring specific artifact research, the methodology supports end-to-end testing where the ground truth is generated via simulated actions, ensuring it is strictly deterministic and free from ambiguity.

Finally, the methodology significantly streamlines cross-tool validation. 
By automating the orchestration of different tools across their respective native environments (e.g., Windows and Linux), it ensures all tools process identical input data. 
This centralized execution reduces the complexity of multi-platform testing, allowing researchers to focus on analyzing discrepancies in tool outputs rather than managing the overhead of disparate experimental setups.


\section{\adare{}}\label{sec:adare}
In this section, we present \adare{}, an implementation of the methodology described above.
We begin by discussing potential implementation strategies and the guiding design principles, 
followed by an over\-view of the system architecture.

\subsection{Implementation Approaches}
We performed a design-level suitability analysis of existing data synthesis frameworks, specifically \fortracepp{}.
While \fortracepp{} excels at generating realistic datasets, it relies on agent-less hypervisor introspection to minimize automation artifacts.
Adapting such a framework for test-driven forensics introduces a fundamental design conflict: effective testing within the guest system necessitates a guest agent, yet the presence of an agent violates the stealth principles central to the \fortracepp{} architecture.~\citep{wolf2024fortracepp}

Additionally, performing all testing after the image creation prevents \statetransitiontesting{}, as it requires tests steps in between user actions.
To resolve these conflicts, we decided to develop a novel framework, \adare{}.
In contrast to the agent-less approach, \adare{} explicitly utilizes an in-guest agent.

\subsection{Design Principles}
Beyond the methodological requirements, the design of \adare{} focused on two additional aspects: reducing barriers to use and maintaining flexibility and extensibility.

\paragraph{Accessibility \& Adoption}
To ensure widespread adoption and practical utility, the framework must prioritize usability and accessibility, minimizing barriers to entry by being easy to install and able to run on commodity hardware. At the same time, it should opportunistically leverage specialized hardware, such as GPUs, to accelerate intensive tasks when available.

The design should provide an intuitive user experience, making both basic and advanced functionalities readily accessible. Furthermore, to foster a collaborative ecosystem and enhance practical application, the framework should provide a catalog of datasets for direct use in experiments, alongside features such as visualization capabilities and mechanisms designed to encourage community engagement and contributions.

\paragraph{Extensibility}
To remain sustainable in the face of rapidly evolving artifacts and forensic tools, the framework must be designed for flexibility and extensibility. 
Its core should remain stable, while selected components, especially the testing, must be easily extendable. 
This ensures that new artifacts, tools, and testing methods can be integrated without major redesigns, allowing the framework to adapt over time and support long-term development.

\subsection{The \adareclient{} Architecture}
\begin{figure}
    \centering
    \includegraphics[width=1\linewidth]{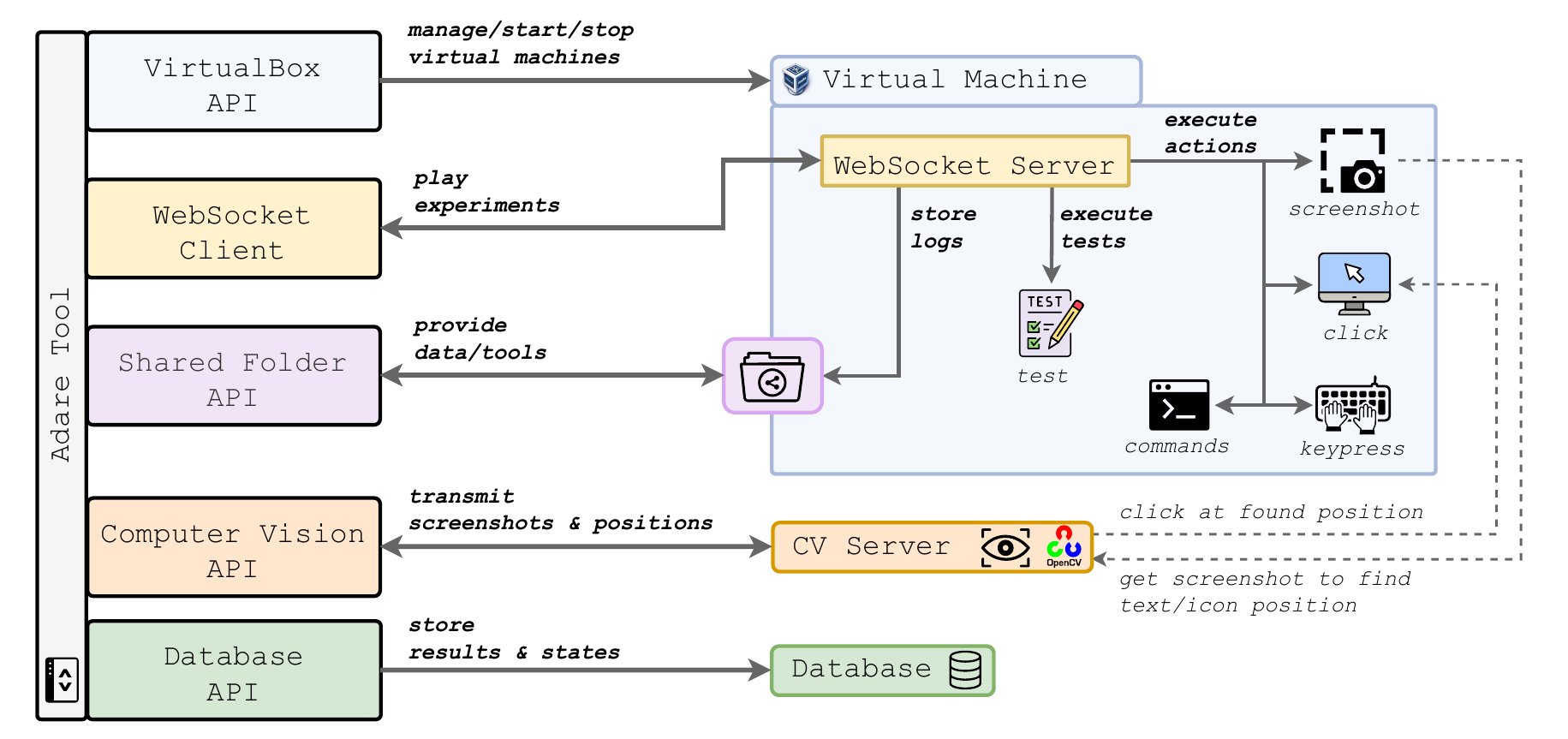}
    \caption{Overview of the \adareclient{} architecture and workflow}
    \label{fig:architecture}
\end{figure}

The \adareclient{} is an open-source command-line application\footnote{\adareclienturl} and serves as the local execution engine of the framework.
The client is designed to run experiments within virtual machines (VMs), enabling both user simulation and testing. 
Its architecture consists of a host-side application and a guest agent, which communicate via a WebSocket connection, as shown in~\cref{fig:architecture}. 
In the following, we cover the basics of \adare{}'s functionality; for an in-depth analysis of specific aspects or guidance on using \adare{}, refer to the tool's documentation\footnote{\adaredocsurl}.

\paragraph{Host-Guest Communication and Execution}
The client supports both VirtualBox and QEMU as backends for managing experimental environments.
Upon starting an experiment, the client launches the defined environment and establishes a connection with the agent running inside the guest. 
To ensure strict reproducibility, every run initializes from a clean VM snapshot.
Once the VM is running, actions are executed sequentially: the host sends a command from the experiment playbook, waits for its completion, processes the result, and then proceeds to the next command.
Because each command can represent either a simulated user interaction or a test, the framework supports \statetransitiontesting{} as described in our methodology, allowing the system state to be tested not only at the end of an experiment but after every step.

\paragraph{Playbooks}
Experiments are defined in playbooks---YAML files that specify both the actions to perform and the tests to verify the outcome.

Actions include GUI interactions (e.g., clicking, typing, scrolling, drag-and-drop) as well as non-GUI operations such as running shell commands.
To verify system state, playbooks invoke test functions from a modular test library. This library supports both unit-level checks (verifying individual values) and integration-level checks (verifying complete artifact structures). Current test categories include standard file checks, format-specific queries (JSON, XML), and SQLite database validation. The framework is also extensible, allowing users to introduce additional test libraries as needed.

Beyond actions and tests, playbooks offer several features that make them easy to create, readable, and flexible enough for complex experiments. 
They support loops and conditional logic to enable structured and adaptive user action and testing flows.
For flexible validation, playbooks provide three types of variables: predefined system variables (such as guest paths), user-defined constants, and dynamic variables populated at runtime.
Dynamic variables are particularly useful for validating runtime specific artifacts. 
For example, a playbook can record the current timestamp immediately after a file deletion and later use that value to verify a corresponding deletion timestamp stored in a \textit{Recycle Bin} artifact.
Playbooks also enable file sharing, making it straightforward to transfer tools or data between the host and the VM.

A minimal playbook demonstrating these concepts is shown in \ref{app:playbook}; a more comprehensive example is available in the project's repository\footnote{\playbookexample}.

\paragraph{Computer Vision for GUI Automation}
To ensure actions are reproducible across different screen resolutions and scaling factors, ADARE minimizes the use of hard-coded screen coordinates. 
Instead, it employs a Computer Vision (CV) API to locate UI elements dynamically. 
By default, this system uses the Scale-Invariant Feature Transform Algorithm for icon detection and PaddleOCR\footnote{\url{https://www.paddleocr.ai/latest/}} for text recognition, allowing the playbook to define GUI actions in human-readable terms rather than pixel positions.

These algorithms were chosen because they provide sufficient accuracy while running efficiently on commodity CPUs. The CV module can also be deployed on GPU-equipped servers and easily extended with machine-learning based approaches.

\subsection{Collaborative Infrastructure: \adareserver{}}
To facilitate the sharing and peer review of experiments, the framework includes the open source \adareserver{}\footnote{\adareserverurl}. 
This platform functions as a central hub where researchers can upload playbooks as well as environment specifications or new test libraries. 
VM images are shared when licensing allows (typically for Linux), while other environments (e.g., Windows) must be reconstructed locally.
Leveraging a Git-based backend, the platform implements a review process for all submissions. 
This moderation step is critical for ensuring scientific soundness and maintaining compliance with legal obligations by verifying resources before they are added to the public repository. 
The web application fosters a federated validation model, where community members can download experiments, replicate them in their own environments, and upload the results to build collective confidence in the findings. 
To further encourage participation, the platform utilizes a badge system to reward active contributors.

\subsection{Enhancing Documentation} 
\adare{} not only facilitates tool testing and validation, but also contributes to the comprehensive and automated documentation of testing activities. It automatically fulfills most documentation requirements for testing results as outlined by the \cite{SWGDE2024testing}.

The author and date of the experiment submission are automatically recorded during the experiment submission to the \adareserver{}. 
The playbooks serve a dual purpose: beyond facilitating test execution, they provide intrinsic, self-contained documentation. Each playbook contains a comprehensive record of the test setup, the sequence of actions performed, and the specific data points being verified, thus documenting the entire testing procedure in a machine- and human-readable format.

The documentation of an experiment's broader purpose and scope is managed through two primary mechanisms. First, these details are explicitly defined by the author during the submission process via the web platform, where they are subject to the platform's established curation and review processes. Second, the purpose and scope are implicitly documented by the content and associated metadata within the playbook itself, which provides a clear technical context for the experiment.


\section{Case Studies}\label{sec:case-study}
In this section, we present \ncasestudies{} case studies that illustrate how \adare{} supports both the digital forensics research community and practitioners.
We selected these case studies to cover complementary evaluation dimensions that reflect common needs in forensic practice and research: (i) \emph{artifact research} (systematically creating, extracting, and validating artifacts across environments and software versions), (ii) \emph{forensic tool testing} (end-to-end regression and differential testing of tools and their outputs), and (iii) \emph{orchestrated cross-environment experiments} (replaying comparable workflows across multiple OS/app configurations).
All related experiments, including their playbooks and environments, are available on GitHub\footnote{\experimentsurl}.

\paragraph{Execution Stability and Portability}
Across all case studies, we further evaluate \adare{} on two practical criteria: (i) \emph{execution stability} (robustness to run-to-run GUI nondeterminism) and (ii) \emph{portability} across host systems.
For stability, we reset to a clean snapshot and executed each experiment at least twice; additionally, we randomly selected one experiment per case study and re-ran it five more times.
We consider a run stable if it completes without human intervention and without nondeterministic automation breakdowns (e.g., pop-ups, focus changes, timing issues, or OCR failed detections).
Across all reruns, we observed no such failures.
For portability, we re-executed one randomly selected experiment per case study on a second host machine with a different CPU; all runs completed successfully and produced consistent outputs according to the experiment-specific oracle.

\subsection{Explorative Artifact Research}
To demonstrate how \adare{} supports explorative artifact research in alignment with the FRED framework, we conducted a case study in which FRED’s methodological principles are instantiated as executable, test-driven experiments within \adare{} to verify the following hypothesis:

\textit{``Deleting a file via Nautilus (GNOME), Dolphin (KDE), or command-line tools (`gio trash', `kioclient5', `trash-put') within the user's home directory and its subdirectories produces identical Trash can artifacts.''}

We created VM environments for Ubuntu 18.04, 20.04, 22.04, and 24.04 (GNOME), as well as Fedora KDE Edition 41 and 42. 
Multiple playbooks were designed, each using a different simulated user actions to achieve deletion but evaluated with the same tests:
\par\noindent
\begin{itemize}
  \itemsep0pt
  \footnotesize
  \item Verify that the original file no longer exists.
  \item Verify that a file with the name of the original exists in \verb!$XDG_DATA_HOME/Trash/files/!.
  \item Verify that a corresponding trashinfo file exists in \verb!$XDG_DATA_HOME/Trash/info/!.
  \item Verify that the info file contains the correct original path and a timestamp matching the deletion time.
\end{itemize}
The results were consistent across all distributions, versions, and deletion methods: all tests passed successfully.

To satisfy the \texttt{Repeat} phase of the FRED methodology, we varied the file path by targeting three different directories within the home directory.
All repeated experiments also passed the tests, further validating the initial hypothesis.
Notably, since \adare{} supports variable parameters (e.g., file paths), additional experiments covering a wide range of edge-case paths can be created with minimal effort.

\subsection{Artifact Regression Testing}\label{sec:artifactregression}
This case study shows how \adare{} can detect changes in artifact behavior across distributions and operating-system updates by testing \texttt{recently-used.xbel}, which records recently accessed resources on systems implementing the freedesktop.org specification~\citep{desktop-bookmark-spec}. 
We executed the same interaction on Ubuntu and Kubuntu Desktop (20.04, 22.04, and 24.04): create a document, open it once, and then validate whether \texttt{recently-used.xbel} exists and whether its access metadata (counts and timestamps) matches the known ground truth.

This approach revealed distribution- and version-specific behavior. 
On Kubuntu, the artifact was absent in 20.04 and 22.04 but present in 24.04 after the same interaction. 
On Ubuntu, the artifact was created in all tested versions, but Ubuntu 20.04 consistently recorded the ``visited'' timestamp as 1969-12-31T23:59:59Z (integer $-1$). 
Later Ubuntu versions recorded timestamps that matched the file access time, and timestamp precision improved over time: Ubuntu 20.04 stored only second-level precision, whereas later versions provided higher-precision values. 
Together, these regressions show that both the presence and the semantics of \texttt{recently-used.xbel} depend on the OS release, and that \adare{} simplifies identifying such changes without the effort of constant manual testing.

\subsection{Tool Validation of PECmd}\label{sec:pecmd}
This case study validates the command-line tool PECmd\footnote{\url{https://github.com/EricZimmerman/PECmd}} by checking whether it correctly parses Windows~11 Prefetch data for a controlled and repeatable execution sequence. 
Instead of evaluating only a final disk image, we applied \statetransitiontesting{} by treating the Prefetch state as a sequence of transitions and validating the tool output after each transition. 
Concretely, we executed a single program (\texttt{regedit}) ten times and, after each execution, invoked PECmd to export its parsed results for the Prefetch directory. 
We then asserted two deterministic expectations at every step: first, that the reported run count increased in lockstep with the iteration number, and second, that the last-run timestamp corresponded to the recorded execution time (within a small tolerance).

To reliably distinguish consecutive executions, we introduced a short delay between runs, because the timestamp recorded in the Prefetch file can lag the externally recorded execution time by a few microseconds and may otherwise lead to ambiguous or colliding timestamps in rapid back-to-back executions. 
Across all iterations, the tests passed, indicating that PECmd produced the expected run-count progression and last-run timestamps for this controlled scenario. 

Overall, this has shown that \adare{} enables tool validation for scenarios like this by verifying, step by step, that tool output matches expected artifact state transitions.

\subsection{Autopsy Tool Regression Testing}
Undocumented changes in forensic tool behavior threaten reproducibility and interpretation, especially for complex GUI-driven tools where workflows are executed manually and internal processing is opaque. To evaluate \adare{}'s ability to render such changes systematically detectable, we performed a regression study of Autopsy\footnote{\url{https://www.autopsy.com/}} across 25 versions, focusing on the \textit{Recent Activity} ingest module and its exported Excel report output.

We used the Windows test image from NIST’s CFReDS repository~\citep{park2016introduction} as a stable input across all runs and defined the expected output using the earliest available version, Autopsy~4.4.0. 
For each version, we executed the same GUI workflow (or a minimally adapted interaction sequence when required by UI changes), while keeping the comparison procedure and divergence criteria identical across all runs.

To capture output divergence consistently, we compared each version’s report against the baseline and categorized deviations as (i) failure to generate the Excel report, (ii) structural changes (e.g., added or removed sheets or columns), (iii) added rows, (iv) removed rows, or (v) cell content changes. Because distinguishing modified rows from added/removed rows is inherently ambiguous in tabular diffs, we applied a simple heuristic: rows with at most two differing cell values were treated as modified; otherwise, they were classified as newly added.

\begin{figure}
    \centering
    \includegraphics[width=1\linewidth]{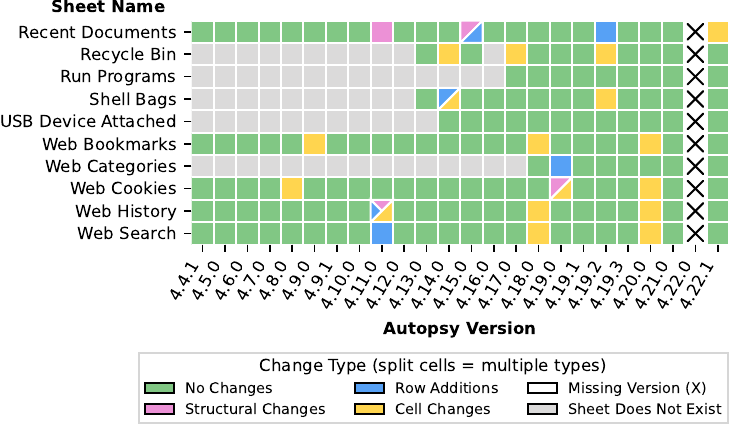}
    \caption{Detected output changes in Autopsy’s \textit{Recent Activity} ingest module across 25 tool versions}
    \label{fig:autopsyversionchanges}
\end{figure}

\Cref{fig:autopsyversionchanges} summarizes the results as a divergence matrix showing detected changes per version. We observed both clear defects and subtle output drift: Autopsy~4.21.0 failed to generate an Excel report entirely, while Autopsy~4.16.0 omitted the \textit{Recycle Bin} sheet.
Across all versions, we identified 16 cell content changes, six row additions, and four structural changes, with no removed rows.

To assess whether these deviations can be also easily identified in other ways, we compared them against Autopsy’s official release notes on GitHub\footnote{\url{https://github.com/sleuthkit/autopsy/releases/}}. Apart from documented browser support additions that align with row additions in \textit{Web History} and \textit{Web Search} in version~4.11.0, and a reported Excel report generation issue in version~4.22.0, we found no entries that explain the remaining observed changes. This suggests that most output deviations occurred without explicit public documentation, making them difficult to anticipate or detect via manual spot checks.

Therefore, this case study demonstrates that \adare{} makes large-scale, multi-version testing of GUI-driven forensic workflows practical by automating repeated execution and systematically exposing output changes that would otherwise be difficult to detect, including changes not reflected in public documentation.

\subsection{Cross-Tool Validation}\label{subsec:cross-tool-validation}
Cross-tool validation is often time-consuming because comparable results require running platform-specific tools in different OS. We demonstrate how \adare{} reduces this overhead by orchestrating three LNK parsers across Windows and Linux VMs: LECmd (Windows), lnkinfo (Linux), and ExifTool (Linux).

We tested three malicious LNK files from VirusTotal: one valid (L1) and two with non-standard appended data (L2, L3). \adare{} ran each parser in its VM and collected outputs on the host. Due to heterogeneous formats, we manually normalized results to align comparable fields.

\begin{table}[ht]
    \centering
    \footnotesize
    \caption{Comparison of LNK parser behavior on three malicious files.
    L1 is structurally valid; L2 and L3 contain non-standard appended data.
    (\ding{51} = success, \ding{55} = failure).}
    \label{tab:lnk-comparison}
    \small 
    \begin{tabular}{llllll}
        \toprule
        \textbf{Tool} & \textbf{Host OS} &  \multicolumn{3}{l}{\textbf{Parsing Success}} & \textbf{Verbosity} \\
         &  & L1 & L2 & L3 & \\
        \midrule
        LECmd     & Windows 11   & \ding{51} & \ding{51} & \ding{51} & High \\
        ExifTool  & Ubuntu 24.04 & \ding{51} & \ding{51} & \ding{51} & Low \\
        lnkinfo   & Ubuntu 24.04 & \ding{51} & \ding{55} & \ding{55} & Medium \\
        \bottomrule
    \end{tabular}
\end{table}

As shown in \cref{tab:lnk-comparison}, all tools parsed the structurally valid LNK file (L1), but differences emerged for files containing appended data (L2, L3): lnkinfo failed due to strict size constraints, whereas LECmd and ExifTool successfully tolerated and bypassed the appended data. The tools also differed in output richness, with LECmd providing the most verbose metadata and ExifTool the least.

Overall, this case study shows that \adare{} makes cross-platform tool comparison repeatable and fast by automating multi-OS execution and consistent output collection, while highlighting that output-format harmonization currently remains a manual step and should be automated in the future to further strengthen comparability.

\section{Limitations}
\adare{} entails several limitations that reflect both inherent tradeoffs of test-driven forensic experimentation and constraints of the current implementation.

First, \adare{} relies on virtualized environments, which inherently limit platform coverage, as not all OS or configurations can be easily or legally virtualized. Virtualization also weakens the connection to physical hardware: external device attachment is hard to automate, virtual USB devices are difficult to emulate, and passthrough may not reproduce bare-metal behavior.

Second, the workflow depends on GUI-driven automation. Although computer vision and OCR reduce reliance on fixed screen coordinates, graphical interfaces evolve across software versions, making playbooks potentially brittle and requiring occasional maintenance. \adare{} mitigates this through modular playbook design and community review, but long-term GUI drift remains an inherent challenge of UI-level testing.

Third, \adare{} uses an in-guest agent rather than an agent-less approach. While this enables fine-grained \statetransitiontesting{} inside the guest system, it can perturb the environment and introduce automation-related traces. For analyses where artifact purity is critical, experiments must explicitly scope and document which agent-induced traces are excluded from interpretation.

Legal and licensing constraints further limit reproducibility. VM images with proprietary OS or applications cannot be redistributed. Instead, \adare{} provides environment specifications, setup instructions, and validation scripts for compliant reconstruction. Additionally, testing multiple versions requires substantial disk space, as older tool and software versions must be retained.

Reproducibility can additionally be threatened by non-determinism and external dependencies, such as background activity, pop-ups, or remote APIs. While \adare{} reduces these effects through controlled execution and explicit assertions, it cannot eliminate all sources of variance. Moreover, the current evaluation focuses primarily on persistent filesystem artifacts; volatile memory and network-centric behaviors are not yet systematically tested.

Finally, as with all specification-based testing, the guarantees provided are bounded by the quality and completeness of the test oracle. Incomplete assertions may fail to detect genuine regressions, while overly permissive tests risk masking subtle changes in artifact behavior.

\section{Conclusion}\label{sec:conclusion}
In this paper, we presented \adare{}, a framework designed to enhance trust and reproducibility in desktop forensics by adapting established software engineering practices. Our work addresses the critical challenges of insufficient tool testing, the dynamic nature of digital artifacts, and the lack of standardized, repeatable experimental procedures in the field. \adare{} enables researchers and practitioners to specify, execute, and share automated experiments within controlled environments, validating outcomes against a known ground truth. The framework comprises (i) a client for running repeatable experiments with computer-vision-guided GUI automation and state-aware validation, and (ii) a web application that supports sharing experiments, assets, and results to strengthen transparency and long-term confidence.

A key methodological contribution is the multi-level testing strategy---unit-, integration-, system-, and regression-like experiments---combined with tests that encode expectations directly into executable playbooks. These playbooks serve a dual purpose: they operationalize artifact and tool expectations as test oracles and simultaneously function as structured, human-readable documentation. By emphasizing \statetransitiontesting{}, ADARE can validate expected system state after each user action, improving causal attribution and making transient or intermediate behaviors testable rather than relying solely on post-mortem disk-image analysis.

Our case studies demonstrate why this approach matters in practice: \adare{} makes otherwise hard-to-observe evolution and drift measurable at scale. On the artifact side, we uncovered distribution- and version-specific divergence in both the presence and semantics of desktop artifacts (including cases where timestamps and precision differed across OS releases). On the tool side, large-scale regression testing revealed undocumented output drifts in a widely used forensic tool across 25 versions, including structural report changes, content changes, and outright failures that were largely not explained by public release notes. Together, these findings illustrate the central value of test-driven forensics: when artifacts and tools evolve silently, executable specifications provide a practical way to detect regressions, preserve interpretability, and support defensible conclusions.

To sustain quality while scaling verification beyond any single lab, \adare{} supports a hybrid governance model: centrally curated submissions (e.g., core environments, playbooks, and test libraries) paired with federated community validation through independent reruns and uploaded results. This structure preserves scientific soundness and legal compliance while enabling broad replication, peer review, and the gradual accumulation of collective confidence. Beyond its primary use cases, the same mechanism also supports related tasks such as automated baseline generation for system-state comparison.

Future work will focus on lowering the cost of experiment authoring, improving robustness against potential UI drift and non-determinism, and enriching result interpretation so that failures yield more actionable explanations rather than only pass/fail outcomes. We also plan to expand test targets beyond persistent filesystem artifacts to include volatile memory and network-derived traces captured at state transitions, and automate cross-tool output normalization and semantic diffing to make multi-tool comparisons reproducible end to end. While this work targets desktop environments, the same paradigm---reproducible user-action generation paired with executable validation---generalizes across digital forensics and provides a practical foundation for continuously verifying forensic claims as platforms evolve.

\bibliography{bibliography}

@article{McKemmish-1999-Forensic-Computing,
  author = {McKemmish, Rodney},
  title = {{What is Forensic Computing?}},
  journal = {{Australian Institute of Criminology Trends and Issues}},
  number = {118},
  year = {1999}
}

@inproceedings{Palmer-2001-dfrws-roadmap,
  title = {{A Road Map for Digital Forensic Research}},
  author = {Palmer, Gary},
  booktitle = {{Proceedings of the first Digital Forensic Research Workshop}},
  pages = {27--30},
  year = {2001}
}

@inbook{Arnes/DF/2-DF-Process,
  author = {Flaglien, Anders O.},
  title = {{Digital Forensics}},
  chapter = {{The Digital Forensics Process}},
  publisher = {{John Wiley \& Sons}},
  year = {2018},
  editor = {Årnes, André}
}

@article{Horsman-FRED,
  title = {{Framework for Reliable Experimental Design (FRED): A research framework to ensure the dependable interpretation of digital data for digital forensics}},
  author = {Horsman, Graeme},
  journal = {{Computers \& Security}},
  volume = {73},
  pages = {294-306},
  year = {2018}
}

@article{Horsman-2019-Tool-Testing,
  title = {Tool testing and reliability issues in the field of digital forensics},
  author = {Horsman, Graeme},
  journal = {{Digital Investigation}},
  volume = {28},
  pages = {163-175},
  year = {2019},
  doi = {10.1016/j.diin.2019.01.009}
}

@book{Martin2011CleanCoder,
  title = {The Clean Coder: A Code of Conduct for Professional Programmers},
  author = {Martin, Robert C.},
  year = {2011},
  publisher = {{Pearson Education, Inc.}}
}

@book{Cohn2010SucceedingWithAgile,
  title = {{Succeeding with Agile: Software Development Using Scrum}},
  author = {Cohn, Mike},
  year = {2010},
  publisher = {{Pearson Education, Inc.}}
}

@misc{amcache-blog-post,
  author = {Eric Zimmerman},
  title = {{binary foray: (Am)cache still rules everything around me (part 2 of 1)}},
  howpublished = {\url{https://binaryforay.blogspot.com /2015/07/amcacheparser-reducing-noise-finding.html}},
  year = {2017}
}

@techreport{lagny-amache,
  title = {{Analysis of the AmCache v2}},
  year = {2019},
  author = {Blanche Lagny},
  institution = {Agence nationale de la sécurité des systèmes d’information (ANSSI)}
}

@article{regression-testing-engstrom,
  author = {Emelie Engström and Per Runeson and Mats Skoglund},
  title = {A systematic review on regression test selection techniques},
  journal = {{Information and Software Technology}},
  volume = {52},
  number = {1},
  pages = {14-30},
  year = {2010},
  doi = {10.1016/j.infsof.2009.07.001}
}

@misc{desktop-bookmark-spec,
  author = {Emmanuele Bassi},
  title = {{desktop-bookmark-spec}},
  howpublished = {\url{https://www.free desktop.org/wiki/Specifications/desktop-bookmark-spec/}},
  year = {2021}
}

@article{agp-grajeda-2018,
  title = {Experience constructing the {A}rtifact {G}enome {P}roject ({AGP}): Managing the domain's knowledge one artifact at a time},
  author = {Cinthya Grajeda and Laura Sanchez and Ibrahim Baggili and Devon Clark and Frank Breitinger},
  journal = {{Digital Investigation}},
  volume = {26},
  pages = {S47-S58},
  year = {2018},
  doi = {10.1016/j.diin.2018.04.021}
}

@article{argus-aardwolf-2025,
  title = {Argus: A new approach for forensic analysis of apps on mobile devices},
  author = {Abdul Boztas and Jeroen {De Jong} and Christos Hadjigeorghiou},
  journal = {Forensic Science International: Digital Investigation},
  volume = {53},
  pages = {301938},
  year = {2025},
  doi = {10.1016/j.fsidi.2025.301938}
}

@article{10yrs-of-cftt-2011,
  author = {James Lyle and Barbara Guttman and Richard Ayers},
  title = {Ten years of computer forensic tool testing},
  year = {2011},
  number = {8},
  journal = {Digital Evidence and Electronic Signature Law Review},
}

@misc{dfir-artifact-museum,
  author = {Andrew Rathburn and Kevin Pagano and Nisarg Suthar and Brian Maloney},
  title = {{GitHub: AndrewRathbun/DFIRArtifactMuseum}},
  howpublished = {\url{https:// github.com/AndrewRathbun/DFIRArtifactMuseum}},
  year = {2022}
}

@misc{df-artifact-repo,
  author = {{The Digital Forensics Artifacts Repository authors}},
  title = {{Digital Forensics Artifacts Repository}},
  howpublished = {\url{https://github.com/ForensicArtifacts/artifacts}},
  year = {2014}
}

@inbook{casey2010forensic-analysis,
  title     = {{Handbook of Digital Forensics and Investigation}},
  editor    = {Eoghan Casey},
  chapter   = {{Forensic Analysis}},
  author    = {Eoghan Casey AND Curtis W. Rose},
  year      = {2010},
  publisher = {{Elsevier Academic Press}}
}

@article{horsman2019raiders,
  author = {Graeme Horsman},
  title = {Raiders of the lost artefacts: Championing the need for digital forensics research},
  journal = {{Forensic Science International: Reports}},
  volume = {1},
  pages = {100003},
  year = {2019},
  doi = {10.1016/j.fsir.2019.100003}
}

@inproceedings{singh2017program,
  title = {{Program Execution Analysis using UserAssist Key in Modern Windows}},
  author = {Singh, Bhupendra and Singh, Upasna},
  booktitle = {{Proceedings of the 14th International Joint Conference on e-Business and Telecommunications (ICETE 2017)}},
  volume = {4},
  pages = {420--429},
  year = {2017},
  organization = {SCITEPRESS -- Science and Technology Publications},
  doi={10.5220/0006416704200429}
}

@inbook{craiger2006le-and-di,
  author    = {J. Philip Craiger AND Jeff Swauger AND Mark Pollitt},
  editor    = {Hossein Bidgoli},
  title     = {{Handbook of Information Security, Volume 2: Information Warfare; Social, Legal, and International Issues; and Security Foundations}},
  chapter   = {{Law Enforcement and Digital Evidence}},
  publisher = {{John Wiley \& Sons, Inc.}},
  year      = {2006}
}

@inbook{craiger2006validation-of-df-tools,
  author    = {J. Philip Craiger AND Jeff Swauger AND Chris Marberry AND Connie Hendricks},
  editor    = {Panagiotis Kanellis AND Evangelos Kiountouzis AND Nicholas Kolokotronis AND Drakoulis Martakos},
  title     = {{Digital Crime and Forensic Science in Cyberspace}},
  chapter   = {{Validation of Digital Forensics Tools}},
  publisher = {{Idea Group Publishing}},
  year      = {2006}
}

@article{OliveiraJr2020-promoting-df-experiments,
  title = {Towards a conceptual model for promoting digital forensics experiments},
  author = {Edson OliveiraJr and Avelino F. Zorzo and Charles Varlei Neu},
  journal = {{Forensic Science International: Digital Investigation}},
  volume = {35},
  pages = {301014},
  year = {2020},
  doi = {10.1016/j.fsidi.2020.301014}
}

@inproceedings{pan2005reproducibility,
  title = {{Reproducibility of Digital Evidence in Forensic Investigations}},
  author = {Lei Pan AND Lynn Batten},
  booktitle = {{Proceedings of the Fifth Digital Forensics Research Workshop (DFRWS)}},
  year = {2005}
}

@book{nap2019rr-in-science,
  author    = {{National Academies of Sciences, Engineering, and Medicine}},
  title     = {{Reproducibility and Replicability in Science}},
  doi       = {10.17226/25303},
  year      = {2019},
  publisher = {{The National Academies Press}},
  address   = {{Washington, DC}}
}

@article{casey2022crowdsourcing,
  author = {Casey, Eoghan and Nguyen, Lam and Mates, Jeffrey and Lalliss, Scott},
  title = {Crowdsourcing forensics: Creating a curated catalog of digital forensic artifacts},
  journal = {{Journal of Forensic Sciences}},
  volume = {67},
  number = {5},
  pages = {1846-1857},
  doi = {10.1111/1556-4029.15053},
  year = {2022}
}

@article{du2021tracegen,
  title = {Trace{G}en: User activity emulation for digital forensic test image generation},
  author = {Xiaoyu Du and Christopher Hargreaves and John Sheppard and Mark Scanlon},
  journal = {{Forensic Science International: Digital Investigation}},
  volume = {38},
  pages = {301133},
  year = {2021},
  doi = {10.1016/j.fsidi.2021.301133}
}

@article{schmidt2023trace-synthesis,
  title = {Improving trace synthesis by utilizing computer vision for user action emulation},
  author = {Lukas Schmidt and Sebastian Kortmann and Thomas Hupperich},
  journal = {{Forensic Science International: Digital Investigation}},
  volume = {45},
  pages = {301557},
  year = {2023},
  doi = {10.1016/j.fsidi.2023.301557}
}

@article{goebel2022fortrace,
  title = {ForTrace - A holistic forensic data set synthesis framework},
  author = {Thomas Göbel and Stephan Maltan and Jan Türr and Harald Baier and Florian Mann},
  journal = {{Forensic Science International: Digital Investigation}},
  volume = {40},
  pages = {301344},
  year = {2022},
  note = {Selected Papers of the Ninth Annual DFRWS Europe Conference},
  doi = {10.1016/j.fsidi.2022.301344}
}

@article{wolf2024fortracepp,
  title = {Hypervisor-based data synthesis: On its potential to tackle the curse of client-side agent remnants in forensic image generation},
  author = {Dennis Wolf and Thomas Göbel and Harald Baier},
  journal = {{Forensic Science International: Digital Investigation}},
  volume = {48},
  pages = {301690},
  year = {2024},
  note = {DFRWS EU 2024 - Selected Papers from the 11th Annual Digital Forensics Research Conference Europe},
  doi = {10.1016/j.fsidi.2023.301690},
}

@InProceedings{goebel2020hystck,
  author = {G{\"o}bel, Thomas and Sch{\"a}fer, Thomas and Hachenberger, Julien and T{\"u}rr, Jan and Baier, Harald},
  editor = {Peterson, Gilbert and Shenoi, Sujeet},
  title = {{A Novel Approach for Generating Synthetic Datasets for Digital Forensics}},
  booktitle = {{Advances in Digital Forensics XVI}},
  year = {2020},
  publisher = {{Springer International Publishing}},
  address = {Cham},
  pages = {73--93}
}

@inproceedings{moch2009forensig,
  author = {Moch, Christian and Freiling, Felix C.},
  title = {The Forensic Image Generator Generator (Forensig2)},
  year = {2009},
  publisher = {IEEE Computer Society},
  address = {USA},
  doi = {10.1109/IMF.2009.8},
  booktitle = {Proceedings of the 2009 Fifth International Conference on IT Security Incident Management and IT Forensics},
  pages = {78–93},
  numpages = {16},
  series = {IMF '09}
}

@article{grajeda2017datasets,
  title = {Availability of datasets for digital forensics – And what is missing},
  author = {Cinthya Grajeda and Frank Breitinger and Ibrahim Baggili},
  journal = {{Digital Investigation}},
  volume = {22},
  pages = {S94-S105},
  year = {2017},
  doi = {10.1016/j.diin.2017.06.004},
}

@article{park2018vmpop,
  title = {{TREDE and VMPOP: Cultivating multi-purpose datasets for digital forensics – A Windows registry corpus as an example}},
  author = {Jungheum Park},
  journal = {Digital Investigation},
  volume = {26},
  pages = {3-18},
  year = {2018},
  doi = {10.1016/j.diin.2018.04.025},
}

@article{scanlon2017eviplant,
  title = {Evi{P}lant: An efficient digital forensic challenge creation, manipulation and distribution solution},
  author = {Mark Scanlon and Xiaoyu Du and David Lillis},
  journal = {{Digital Investigation}},
  volume = {20},
  pages = {S29-S36},
  year = {2017},
  note = {DFRWS 2017 Europe},
  doi = {10.1016/j.diin.2017.01.010}
}

@inproceedings{sremack2007gap,
  title = {{The Gap between Theory and Practice in Digital Forensics}},
  author = {Sremack, Joseph C.},
  booktitle = {{Annual ADFSL Conference on Digital Forensics, Security and Law (ADFSL)}},
  year = {2007}
}

@article{tully2020quality-standards,
  title = {Quality standards for digital forensics: Learning from experience in {E}ngland \& {W}ales},
  author = {Gillian Tully and Neil Cohen and David Compton and Gareth Davies and Roy Isbell and Tim Watson},
  journal = {{Forensic Science International: Digital Investigation}},
  volume = {32},
  pages = {200905},
  year = {2020},
  doi = {10.1016/j.fsidi.2020.200905}
}

@article{stoykova2023reliability,
  title = {Reliability validation enabling framework ({RVEF}) for digital forensics in criminal investigations},
  author = {Radina Stoykova and Katrin Franke},
  journal = {{Forensic Science International: Digital Investigation}},
  volume = {45},
  pages = {301554},
  year = {2023},
  doi = {10.1016/j.fsidi.2023.301554}
}

@article{horsman2019derds,
  title = {Formalising investigative decision making in digital forensics: {P}roposing the {D}igital {E}vidence {R}eporting and {D}ecision {S}upport ({DERDS}) framework},
  author = {Graeme Horsman},
  journal = {{Digital Investigation}},
  volume = {28},
  pages = {146-151},
  year = {2019},
  doi = {10.1016/j.diin.2019.01.007}
}

@article{breitinger2023sharing,
  title = {Sharing datasets for digital forensic: A novel taxonomy and legal concerns},
  author = {Frank Breitinger and Alexandre Jotterand},
  journal = {{Forensic Science International: Digital Investigation}},
  volume = {45},
  pages = {301562},
  year = {2023},
  doi = {10.1016/j.fsidi.2023.301562}
}

@article{horsman2025traces,
  author = {Graeme Horsman},
  title = {Understanding and comparing digital traces},
  journal = {{Australian Journal of Forensic Sciences}},
  volume = {57},
  number = {4},
  pages = {481--491},
  year = {2025},
  publisher = {Taylor \& Francis},
  doi = {10.1080/00450618.2024.2381535},
}

@article{pan2009performance,
  title = {Robust performance testing for digital forensic tools},
  author = {Lei Pan and Lynn M. Batten},
  journal = {{Digital Investigation}},
  volume = {6},
  number = {1},
  pages = {71-81},
  year = {2009},
  doi = {10.1016/j.diin.2009.02.003}
}

@misc{SWGDE2024testing,
  title = {{Minimum Requirements for Testing Tools Used in Digital and Multimedia Forensics, Version: 2.1}},
  year = {2024},
  author = {{Scientific Working Group on Digital Evidence}}
}

@misc{flandrin2014dfts,
  title = {{Evaluating Digital Forensic Tools (DFTs)}},
  year = {2014},
  author = {Flavien Flandrin AND William J. Buchanan AND Richard Macfarlane AND Bruce Ramsay AND Adrian Smales}
}

@article{horsman2018yourhonour,
  title = {“I couldn't find it your honour, it mustn't be there!” – Tool errors, tool limitations and user error in digital forensics},
  journal = {{Science \& Justice}},
  author = {Graeme Horsman},
  volume = {58},
  number = {6},
  pages = {433-440},
  year = {2018},
  doi = {10.1016/j.scijus.2018.04.001}
}

@article{guo2009validation,
  title = {Validation and verification of computer forensic software tools—Searching Function},
  author = {Yinghua Guo and Jill Slay and Jason Beckett},
  journal = {{Digital Investigation}},
  volume = {6},
  pages = {S12-S22},
  year = {2009},
  note = {The Proceedings of the Ninth Annual DFRWS Conference},
  doi = {10.1016/j.diin.2009.06.015}
}

@inproceedings{wilsdon2006blackbox,
  title = {{Validation of forensic computing software utilizing Black Box testing techniquestesting techniques}},
  author = {Tom Wilsdon AND Jill Slay},
  booktitle = {{Proceedings of the 4th Australian Digital Forensics Conference}},
  year = {2006}
}

@article{lyle2010errorrate,
  title = {If error rate is such a simple concept, why don’t I have one for my forensic tool yet?},
  author = {James R. Lyle},
  journal = {{Digital Investigation}},
  volume = {7},
  pages = {S135-S139},
  year = {2010},
  note = {The Proceedings of the Tenth Annual DFRWS Conference},
  doi = {10.1016/j.diin.2010.05.017}
}

@article{bhat2021trusted,
  title = {Can computer forensic tools be trusted in digital investigations?},
  author = {Wasim Ahmad Bhat and Ali AlZahrani and Mohamad Ahtisham Wani},
  journal = {Science \& Justice},
  volume = {61},
  number = {2},
  pages = {198-203},
  year = {2021},
  issn = {1355-0306},
  doi = {10.1016/j.scijus.2020.10.002}
}

@inproceedings{wundram2013antiforensics,
  author = {Wundram, Martin and Freiling, Felix C. and Moch, Christian},
  booktitle = {{2013 Seventh International Conference on IT Security Incident Management and IT Forensics}},
  title = {{Anti-forensics: The Next Step in Digital Forensics Tool Testing}},
  year = {2013},
  pages = {83-97},
  doi = {10.1109/IMF.2013.17}
}

@article{Brunty2023,
  author = {Josh Brunty},
  title = {Validation of forensic tools and methods: A primer for the forensic science practitioner},
  journal = {Wiley Interdisciplinary Reviews: Forensic Science},
  year = {2023},
  volume = {5},
  number = {2},
  pages = {e1474},
  doi = {10.1002/wfs2.1474},
}

@article{spichinger2025evidencemeaning,
  title = {Preserving meaning of evidence from evolving systems},
  author = {Hannes Spichiger and Frank Adelstein},
  journal = {{Forensic Science International: Digital Investigation}},
  volume = {52},
  pages = {301867},
  year = {2025},
  note = {DFRWS EU 2025 - Selected Papers from the 12th Annual Digital Forensics Research Conference Europe},
  doi = {10.1016/j.fsidi.2025.301867}
}

@article{hargreaves2025solve,
  title={SOLVE-IT: A proposed digital forensic knowledge base inspired by MITRE ATT\&CK},
  author={Hargreaves, Christopher and van Beek, Harm and Casey, Eoghan},
  journal={Forensic Science International: Digital Investigation},
  volume={52},
  pages={301864},
  year={2025},
  publisher={Elsevier}
}

@article{hargreaves2024abstract,
  title={An abstract model for digital forensic analysis tools-a foundation for systematic error mitigation analysis},
  author={Hargreaves, Christopher and Nelson, Alex and Casey, Eoghan},
  journal={Forensic Science International: Digital Investigation},
  volume={48},
  pages={301679},
  year={2024},
  publisher={Elsevier}
}

@article{rzepka2025scenario,
  title={A scenario-based quality assessment of memory acquisition tools and its investigative implications},
  author={Rzepka, Lisa and Ottmann, Jenny and Stoykova, Radina and Freiling, Felix and Baier, Harald},
  journal={Forensic Science International: Digital Investigation},
  volume={52},
  pages={301868},
  year={2025},
  publisher={Elsevier}
}

@article{voigt2025metrics,
  title={A metrics-based look at disk images: Insights and applications},
  author={Voigt, Lena L and Freiling, Felix and Hargreaves, Christopher},
  journal={Forensic Science International: Digital Investigation},
  volume={52},
  pages={301874},
  year={2025},
  publisher={Elsevier}
}

@article{park2016introduction,
  title={Introduction to CFTT and CFReDS projects at NIST},
  author={Park, Jungheum and Lyle, James R and Guttman, Barbara},
  year={2016},
  publisher={Jungheum Park, James R. Lyle, Barbara Guttman}
}

@phdthesis{carrier2006hypothesis,
  title={A hypothesis-based approach to digital forensic investigations},
  author={Carrier, Brian D},
  year={2006},
  school={Purdue University}
}

@article{garfinkel2012general,
  title={A general strategy for differential forensic analysis},
  author={Garfinkel, Simson and Nelson, Alex J and Young, Joel},
  journal={Digital Investigation},
  volume={9},
  pages={S50--S59},
  year={2012},
  publisher={Elsevier}
}

@article{moreau2023containers,
  title={Containers for computational reproducibility},
  author={Moreau, David and Wiebels, Kristina and Boettiger, Carl},
  journal={Nature Reviews Methods Primers},
  volume={3},
  number={1},
  pages={50},
  year={2023},
  publisher={Nature Publishing Group UK London}
}

\appendix

\section{LNK Files}
The LNK files used in \cref{subsec:cross-tool-validation} were obtained from VirusTotal. The hashes are:
\begin{itemize}
    \itemsep0pt
    \item LNK-1: \texttt{\footnotesize acbc775087da23725c3d783311d5f5083c93658de392c1 7994a9151447ac2b63}
    \item LNK-2: \texttt{\footnotesize 1b75f70c226c9ada8e79c3fdd987277b0199928800c51e 5a1e55ff01246701db}
    \item LNK-3: \texttt{\footnotesize 1b598c7c35f00d2c940dfd3745bd9e5d036df781d391b8 f3603a2969c666761b}
\end{itemize}

Software versions used for the experiment are: LECmd v.1.5.1, lnkinfo v.20181227 and ExifTool v.12.76

\section{Minimal Playbook Example}\label{app:playbook}
\begin{lstlisting}[language=yaml]
variables:
  filename:
    type: string
    value: "secret.txt"
  filepath:
    type: path
    value: "{{ adare_user_documents }}/{{ filename }}"
  trashbin:
    type: path
    value: "{{ adare_user_home }}/.local/share/Trash"

tests:
  - name: file_in_trash
    function: file_exists
    parameter:
      dst: "{{ trashbin }}/files/{{ filename }}"
  - name: trashinfo_exists
    function: file_exists
    parameter:
      dst: "{{ trashbin }}/info/{{ filename }}.trashinfo"

actions:
  - command:
      command: "echo secret > {{ filepath }}"
      shell: true
  - click:
      target:
        image: "nautilus_taskbar.png"
  - click:
      target:
        text: "Documents"
  - click:
      type: "right"
      target:
        text: "{{ filename }}"
  - click:
      target:
        text: "Move to Trash"
  - test: file_in_trash
  - test: trashinfo_exists
\end{lstlisting}

\section{Notes}
An example playbook that verifies the \textit{Recycle Bin} behavior on an Ubuntu system and demonstrates many of the aforementioned playbook features---along with a screen-recorded video of the experiment---is available at \adaredemourl{}, along with a second example that verifies the \textit{Recycle Bin} behavior in Windows 11.

\end{document}